\documentclass[aps,prd,twocolumn,showpacs,superscriptaddress,groupedaddress,nofootinbib]{revtex4-1}
\pdfoutput=1

\usepackage{nicefrac}
\usepackage[latin1]{inputenc}
\usepackage{graphicx}
\usepackage{amssymb,amsmath}

\usepackage{mdwlist}
\usepackage[bookmarks=true,bookmarksopen=true,pdfhighlight=/I,pdfpagemode=UseOutlines,linktocpage,plainpages=false]{hyperref}
\usepackage{amsfonts}
\usepackage[usenames,dvipsnames]{color}
\usepackage[all]{xy}
\hypersetup{colorlinks=true,urlcolor=Sepia,linkcolor=BlueViolet,citecolor=OliveGreen}
\usepackage{IEEEtrantools}

\newcommand{\be}{\begin{equation}}
\newcommand{\ee}{\end{equation}}
\newcommand{\bea}{\begin{IEEEeqnarray}{rCl}}
\newcommand{\eea}{\end{IEEEeqnarray}}
\newcommand{\nn}{\nonumber\\}
\newcommand{\ba}{\begin{array}}
\newcommand{\ea}{\end{array}}

\newcommand{\w}{\omega}

\newcommand{\rfv}{|0\rangle_{f}}
\newcommand{\lfv}{_{f}\langle 0 |}
\newcommand{\rmv}{| 0 \rangle}
\newcommand{\lmv}{\langle 0 |}
\newcommand{\bpsi}{\bar{\psi}}

\newcommand{\vk}{\vec{k}}

\newcommand{\vx}{\vec{x}}

\newcommand{\e}{\eta}

\newcommand{\lag}{\mathcal L}

\newcommand{\st}{\sin \theta}
\newcommand{\sst}{\sin^2 \theta}

\newcommand{\ct}{\cos \theta}
\newcommand{\cct}{\cos^2 \theta}

\newcommand{\diracpartial}{\partial\!\!\!/}

\newcommand{\dg}{\dagger}

\newcommand{\field}[1]{\mathbb{#1}}

\newcommand{\hilbert}[1]{$\mathfrak{#1}$}
\newcommand{\ket}[1]{| #1 \rangle}
\newcommand{\bra}[1]{\langle #1 |}

\definecolor{lightGray}{RGB}{220,220,220}
\newcommand{\slot}{\textcolor{lightGray}{\blacksquare}}
\definecolor{MyDarkBlue}{rgb}{0,0.08,0.45}

\definecolor{MyDarkGray}{RGB}{140,140,140}

\newcommand{\bes}{\begin{equation}\begin{split}}
\newcommand{\ees}{\end{split}\end{equation}}

\definecolor{lightGray}{RGB}{220,220,220}
\definecolor{MyDarkGray}{RGB}{140,140,140}

\definecolor{MyDarkBlue}{rgb}{0,0.08,0.45}

\newcommand{\fvev}[1]{\lmv \because #1 \because \rmv}
\newcommand{\fnormal}[1]{\because #1 \because}

\newcommand{\pressure}{\mathsf{P}}

\newlength{\eqboxstorage}

\begin{document}

\widetext
\title{Dark Matter and Dark Energy via Non-Perturbative (Flavour) Vacua}
\author{Walter Tarantino}\email{walter.tarantino@kcl.ac.uk}
\affiliation{King's College London, Department of Physics, Strand, London WC2R 2LS, UK.}
\begin{abstract}
A non-perturbative field theoretical approach to flavour physics (Blasone-Vitiello formalism) has been shown to imply a highly non-trivial vacuum state.
Although still far from representing a
satisfactory framework for a coherent and complete characterization of flavour states, in recent
years the formalism has received attention for its possible implications at cosmological scales.
In a previous work, we implemented the approach on a simple supersymmetric model (free Wess-Zumino), with flavour mixing,
which was regarded as a model for free neutrinos and sneutrinos.
The resulting effective vacuum (called \textit{flavour vacuum}) was found to be characterized by a strong SUSY breaking.
In this paper we explore the phenomenology of the model and we argue that the flavour vacuum is
a consistent source for both Dark Energy
(thanks to the bosonic sector of the model) and Dark Matter (via the fermionic one).
Quite remarkably, besides the parameters connected with neutrino physics, in this model
no other parameters have been introduced,
possibly leading to a predictive theory of Dark Energy/Matter. Despite its oversimplification,
such a toy model already seems capable to shed some light on the observed energy hierarchy between neutrino physics,
Dark Energy and Dark Matter.
Furthermore, we move a step forth in the construction of a more realistic theory, by presenting a novel approach for calculating relevant quantities
and hence extending some results to interactive theories, in a completely non-perturbative way.
\end{abstract}

\maketitle

\section*{Introduction}

Neutrino Flavour Oscillation is nowadays a fairly well-established fact, thanks to a wide range of experimental evidences \cite{Strumia:2006db}.
A simple quantum mechanical model (based on the work of Pontecorvo,  Maki, Nagawa, and Sakata \cite{Pontecorvo:1957qd,Pontecorvo:1967fh,Gribov:1968kq,Maki:1962mu})
is commonly considered as sufficient for accounting for experimental data.
However, this hides non-trivial difficulties in the formulation of flavour oscillations in a Quantum Field Theoretical (QFT) framework \cite{Beuthe:2001rc}.
Flavour states indeed do not represent correct asymptotic states (by definition, since their oscillating behaviour),
which are required in the usual perturbative approach to QFT (in the Lehmann, Symanzik and Zimmermann scheme).\\
More than a decade ago, a non-perturbative approach for building flavour states was suggested by Blasone, Vitiello and coworkers
(\textit{BV formalism} for flavour physics) \cite{Blasone:1995zc}. 
A first version was proposed in 1995 \cite{Blasone:1995zc}, but some inconsistencies in the
derivation of oscillation formulae were noticed \cite{Fujii:1998xa,Blasone:1999jb,Fujii:2001zv,Blasone:2001sr} shortly after;
a revisited version, in which these discrepancies were clarified and removed,
was suggested and developed \cite{Blasone:1998hf,Blasone:2001du,Ji:2001yd,Ji:2002tx,Blasone:2002jv} later on \cite{Beuthe:2001rc}. However,
for some aspects, the formalism remains controversial and its physical relevance is still matter of debate \cite{Giunti:2003dg,Giunti:2006fr,Blasone:2005ae}.\\
The approach correctly reduces to the common quantum mechanical approach in the small (neutrinos) mass limit, but leads to
corrections to those formulas that are currently beyond experimental sensitivity \cite{Beuthe:2001rc,Capolupo:2004av}.
However, perhaps the most interesting feature of BV formalism is the non-trivial vacuum
(called \textit{flavour vacuum}) implied by the theory.
Such a \textit{flavour vacuum} (which can be regarded as a \textit{vacuum condensate})
represents the physical state with no (flavour) particles in it. Despite being merely an ``empty'' state,
the flavour vacuum is characterized by a rich structure
revealed by the non-vanishing expectation value of the stress-energy tensor and the related equation of state.
Within BV formalism one is able to fully describe it at a non-perturbative level,
and its features depend on the specific model considered
(spin, interactions, number of particles characterized by flavour mixing, etc.).\\
In a series of papers it was suggested that the \textit{flavour vacuum} might
behave as a source of Dark Energy
\cite{Blasone:2004hr,Blasone:2004yh,Capolupo:2006et,Capolupo:2006re,Capolupo:2008rz,Capolupo:2007hy,Blasone:2007iq,Blasone:2008rx,Blasone:2007jm,Blasone:2006ch}.
Recently, it has been shown that in a simple supersymmetric (Wess-Zumino) model with flavour mixing,
in which two Majorana fields,
two scalars and two pseudo-scalars were present (a simple model for neutrinos and sneutrinos),
the \textit{flavour vacuum} was actually characterized by a strong Supersymmetry breaking
\cite{Mavromatos:2010ni,Capolupo:2010ek}.
In the present work we argue that this breaking is the origin of an interesting phenomenology,
that might shed some lights both on the Dark Energy and the Dark Matter problem.
More precisely, in the supersymmetric context of \cite{Mavromatos:2010ni,Capolupo:2010ek},
the \textit{flavour vacuum} can be thought as made of two different fluids that fill in all universe:
a first one related to the bosonic sector of the model, and a second induced by the fermionic one.
The former is characterized by negative pressure, equal in modulus to its energy density (acting as a source of Dark Energy).
The latter is characterized by zero pressure, giving rise to a source of Cold Dark Matter.

The first part of the paper (Section \ref{section:BVformalism})
will be dedicated to review BV formalism, complementing the original literature with a discussion on non-perturbative theories and Fock spaces.

In the second part of the paper (Section \ref{section:Phenomenology}) we shall explore in details the phenomenology of the model studied in \cite{Mavromatos:2010ni,Capolupo:2010ek}.
We shall clarify why the \textit{flavour vacuum}
is a good Dark Energy and Dark Matter candidate, with emphasis on this latter.
More importantly, we will explain how to relate all parameters of the model to observational data.
This is a quite important aspect of the approach:
the model introduces very few parameters which are all related to neutrino physics.
Hopefully, more realistic models will not rely on uncontrolled free parameters, leading to a truly
falsifiable theory for both Dark Energy and Dark Matter.
A first encouraging result in this direction comes already from the simple model here studied:
this is indeed fairly consistent with a choice of the parameters
modeled on real world data, as we shall see in the relevant section.

A first step towards more realistic models will be moved in the last part of this work
(Section \ref{section:Interactions}).
We will present a novel method for calculating relevant quantities, specifically
thought for analyzing the features of the \textit{flavour vacuum}, which
might enable us to study interactive theories completely at a non-perturbative level.
In particular, we shall show how, under reasonable assumptions, the method can
discriminate which interactions preserve the behaviour
of the condensate as Dark Energy/Matter source.

\section{BV Formalism}\label{section:BVformalism}

Neutrino oscillations can be described in a non-relativistic
quantum mechanical framework by constructing particle states,
labelled by a \textit{flavour number}, that are not eigenstates of the Hamiltonian.
In its simplest formulation for two distinct flavours, the Pontecorvo model,
flavour states are constructed as follow \cite{Bilenky:2010zza}:
\bea\label{pontecorvostates}
\ket{\nu_A}&=&\ct \ket{\nu_1} +\st \ket{\nu_2}\nn
\ket{\nu_B}&=&-\st \ket{\nu_1} +\ct \ket{\nu_2}
\eea	
where $\ket{\nu_1}$ and $\ket{\nu_2}$ are massive eigenstate of the free Hamiltonian (particles with well defined mass $m_i$, with $i=1,2$), and from which
\bea\label{pontecorvoformula1}
\wp_{A\rightarrow B}&=&|\langle \nu_B\ket{\nu_A(t)}|^2=|\bra{\nu_B}e^{-iH t}\ket{\nu_A}|^2\nn
									&=&\sst \sin^2\left(\frac{\w_1(k) t-\w_2(k) t}{2}\right)
\eea
with $\w_i(k)\equiv \sqrt{\vk^2+m_i^2}$,
describing the non-vanishing probability of a flavoured particle with momentum $\vk$ to be created with a certain flavour ($A$)
and be detected later on with a different flavour ($B$).

The form of the transformation between flavoured and massive particles (\ref{pontecorvostates})
is reflected in the relativistic field formalism by the relation
\bea\label{flavourandmassive}
\nu_A(x)&=& \nu_1 (x) \ct+\nu_2 (x) \st\nn
\nu_B(x)&=& -\nu_1 (x) \st+\nu_2 (x) \ct
\eea
that connects flavour fields $\nu_A$, $\nu_B$ with massive ones $\nu_1$, $\nu_2$. Such a relation is connected with the
linearization of the following Lagrangian for free spin-\nicefrac{1}{2} fields
\begin{multline}\label{lagflavour}
\lag=i\bar{\nu}_A(x)\diracpartial \nu_A(x)+i\bar{\nu}_B(x)\diracpartial \nu_B(x)+\\
      -m_A \bar{\nu}_A(x) \nu_A(x)-m_B \bar{\nu}_B(x) \nu_B(x)+\\
      -m_{AB}\left(\bar{\nu}_A(x) \nu_B(x)+\bar{\nu}_B(x) \nu_A(x)\right)
\end{multline}
which becomes
\begin{multline}\label{lagmassive}
\lag=i\bar{\nu}_1(x)\diracpartial \nu_1(x)+i\bar{\nu}_2(x)\diracpartial \nu_2(x)+\\
      -m_1 \bar{\nu}_1(x) \nu_1(x)-m_2 \bar{\nu}_2(x) \nu_2(x)
\end{multline}
when
\bea
m_A&=&m_1\cct+m_2 \sst\nn
m_B&=&m_1 \sst+m_2 \cct\nn
m_{AB}&=& (m_2-m_1) \st \ct.
\eea

However, in the field-theoretical framework the decomposition of the fields (\ref{flavourandmassive})
into ladder operators associated with flavour particle states is highly non-trivial \cite{Beuthe:2001rc}.
It has been shown, indeed, that states defined as the relativistic equivalent of (\ref{pontecorvostates}),
for which $\ket{\nu_{1,2}}$ belongs to the mass-$m_{1,2}$ irreducible representation of the Poincar\'e group,
are \textit{not} eigenstates of the
flavour charge operators \cite{Blasone:2006jx,Blasone:2008ii}, which for the theory (\ref{lagflavour}) read \cite{Blasone:2008ii}
\bea\label{flavourcharges}
Q_A(t)&=&\int d\vx\; \nu^{\dg}_A(x)\nu_A(x),\nn
Q_B(t)&=&\int d\vx\; \nu^{\dg}_B(x)\nu_B(x).
\eea

BV formalism compensate for this \cite{Blasone:1995zc,Hannabuss:2000hy,Ji:2002tx}, by defining appropriate flavour eigenstates via the action of
a certain operator $G_\theta$ on massive states:
\begin{multline}
\ket{\vec{k}_1,f_1;\vec{k}_2,f_2;\vec{k}_3,f_3;...}\equiv \\ G^\dg_\theta\ket{\vec{k}_1,m_{(1)};\vec{k}_2,m_{(2)};\vec{k}_3,m_{(3)};...}
\end{multline}
where $\ket{\vec{k}_1,f_1;\vec{k}_2,f_2;\vec{k}_3,f_3;...}$ denotes a state with different flavour particles described by their momenta $\vec{k}_i$ and their flavours $f_i=A,B$,
whereas $\ket{\vec{k}_1,m_{(1)};\vec{k}_2,m_{(2)};\vec{k}_3,m_{(3)};...}$ denotes a state with different massive particles described by their momenta $\vec{k}_i$ and their masses $m_{(i)}=m_1,m_2$, defined from the linearized
theory (\ref{lagmassive}).
The operator $G_\theta$ is defined by the equations:
\bea\label{G1WT}
\nu_A(x)&=& G^{-1}_{\theta} \nu_1 (x) G_{\theta}\nn
\nu_B(x)&=& G^{-1}_{\theta} \nu_2 (x) G_{\theta}
\eea
and its explicit form depends on the specific theory considered. For the theory (\ref{lagflavour}),
$G_{\theta}$ is written as \cite{Blasone:1995zc}
\be
G_{\theta}(t)=e^{\frac{\theta}{2}\int d\vx \left( \nu^\dg_1(x)\nu_1(x)-\nu^\dg_1(x)\nu_2(x) \right)}.
\ee
Among all flavour states defined in the BV approach, the one called \textit{flavour vacuum} and defined by
\be\label{fvdef}
\rfv\equiv G^\dg_\theta \rmv
\ee
plays a special role, since it represents the physical vacuum.
In this context, by \textit{physical vacuum} we mean the
state that represents the physical empty state, \textit{i.e.} with no particle in it.
Since only particles with well defined flavour, rather than mass (i.e. Hamiltonian eigenstates),
can be created/detected, one expects the physical vacuum to be represented by the state that
counts no flavour particles in it. It can be shown that this is state is $\rfv$, rather than $\rmv$ \cite{Blasone:1995zc}.\footnote{In
short: flavour states can be built by means of specific creation/annihilation operators
for flavour particles; it can be shown that
the only state that is annihilated by all annihilation operators is the flavour vacuum
defined via (\ref{fvdef}).}

Furthermore, it has been proven that \textit{all} flavour states $\ket{\vec{k}_1,f_1;\vec{k}_2,f_2;\vec{k}_3,f_3;...}$ are \textit{orthogonal}
to each massive state $\ket{\vec{k}_1,m_{(1)};\vec{k}_2,m_{(2)};\vec{k}_3,m_{(3)};...}$, and therefore $\langle 0 \rfv=0$ follows as a particular case.
This result enables us to talk of a \textit{Fock space for flavour states}, in opposition
of the usual Fock space, whose basis is given by $\{\ket{\vec{k}_1,m_{(1)};\vec{k}_2,m_{(2)};\vec{k}_3,m_{(3)};...\vec{k}_n,m_{(n)}} \mid \forall n\in \field{N} \}$.

The orthogonality of the two spaces is not very surprising
if one regards BV formalism as a non-perturbative approach to the
interactive theory defined by (\ref{lagflavour}).

In the Second-quantization framework, the Hilbert space representing physical states is defined by vectors in the number occupation representation:
assuming that a single-particle state can be classified by a discrete set of
states labeled by the index $i=1,2,3,...$, a vector representing a many-particle state can
be identified by the number $n_i$ of particles occupying the $i$-th state and it is
denoted with $\ket{n_1,n_2,n_3,...}$;
the Hilbert space of physical states \hilbert{H} is therefore defined as the vector space generated
by the basis $\{\ket{n_1,n_2,n_3,...}\}$.
For bosons $n_i\in \field{N}_0$, whereas for fermions $n_i=0,1$.
It can be shown that in both cases the set $\{\ket{n_1,n_2,n_3,...}\}$ is \textit{uncountable} and therefore \hilbert{H} is non-separable \cite{Umezawa:1982nv}.
In particle physics a separable subset of \hilbert{H}, a Fock space that we will denote with \hilbert{F_0}, is usually considered \cite{Streater:1989vi}.
\hilbert{F_0} carries an irreducible representation of the Poincar\'e group and particle states belonging to it
have well-defined mass and spin.
Such a subset is spanned by the \textit{countable} basis
of all states with an arbitrary, yet finite in total, number of free particles.
Although this basis does not fully describe \hilbert{H} (for instance the vector $\ket{1,1,1,...}$ which counts
an infinite number of particles is not included), it is sufficient for accounting for scattering processes at a \textit{perturbative} level.
In the usual perturbation theory all \textit{interactive} processes are indeed approximated by means of a
superposition of a finite number of \textit{free} particle states.
This is quite clear in the functional formalism, when Feynman diagrams are considered.
In a simplified picture of this framework, a scattering process is represented
by a graph with a certain number of external legs (incoming and outgoing particles).
The total number of internal lines
is connected to the precision of the approximation used in the perturbative expansion:
the higher the order of the perturbation, the higher the number of vertices,
and therefore the higher the number of internal lines involved. Each line can be na\"ively interpreted as
a single free particle state, which is emitted in the starting vertex and then absorbed
in the ending one. At each order in perturbation theory, a finite number of free
particle states enters in the description of the scattering process.
However, under the assumption that the perturbative series converges,
its limit would be described by an \textit{infinite} number of
lines/one-free-particle-states. In bra-ket formalism such a limit state would therefore be represented
by a vector of \hilbert{H}, the space of \textit{all} physical states, but not of \hilbert{F_0},
the space of states with \textit{finite} number of free particles.
With this example we want to remark the non-trivial difference existing from a \textit{free} theory and an \textit{interactive} one:
we can express interactive processes in terms of free states (which have no direct physical meaning or interpretation, being just a basis
in which we choose to express our process) but only in a weakly-interactive/perturbed framework. A full non-perturbative treatment for interactive particle states requires
subspaces of \hilbert{H}, that are orthogonal to the Fock space of free states \hilbert{F_0} \cite{Haag:1955ev}.

Coming back to our case, a Lagrangian with flavour mixing, such as (\ref{lagflavour}), can be regarded as
an interactive theory, thanks to its non-diagonal terms.
Flavour particle states defined \textit{\`a la} BV form a Fock space \hilbert{F_f} that is therefore orthogonal to \hilbert{F_0}.
In other words, we could express flavour states in a perturbative way by means of \hilbert{F_0}; however, BV formalism enables us to construct
flavour states in a completely non-perturbative manner, and therefore it requires states that are part of \hilbert{H} but not of \hilbert{F_0}.

Different Fock spaces are ordinarily used in QFT on curved backgrounds
and other contexts \cite{Birrell:1982ix,calzetta2008nonequilibrium}.
In the former, for instance, one identifies Fock spaces for physical states in flat regions.
However, these Fock spaces do not coincide (they are different/orthogonal subset of \hilbert{H}) if the regions are not connected and curved regions exist in between.
As a consequence, the vacuum defined by an observer in a certain region is not necessarily described by the state with no particles by an observer in another region.
The \textit{particle creation} phenomenon may occur: a state that is empty for an observer can actually contain particles according with a different observer.
This mechanism characterizes both of the two main results of QFT in curved spacetime: the Unruh effect \cite{Unruh:1976db} and the Hawking radiation \cite{Hawking:1974sw}.

In a formal analogy, BV formalism introduces a ground state, called \textit{flavour vacuum}, which is not as trivial as the ground state for the free theory.
As already said, since it is the state in which no flavour particles are present, it correctly represents the \textit{physical vacuum}.
Even though it is \textit{empty}, it is characterized by a non-zero expectation value of the stress-energy tensor
$\lfv T_{\mu\nu}\rfv$, whose effects must be gravitationally testable. This is true as long as we fix as zero-point of our theory
the usual vacuum $\rmv$ for the free theory and belonging to \hilbert{F_0},
or, in other words, we consider the usual \textit{normal ordering} $\lfv :T_{\mu\nu}:\rfv\equiv \lfv T_{\mu\nu}\rfv-\lmv T_{\mu\nu}\rmv$,
which is valid in perturbation theory as well as in this non-perturbative approach.
One commonly refers to the flavour vacuum as a \textit{condensate} for the following reason: once expressed in terms of particles with well defined mass (eigenstates of the Hamiltonian),
the flavour vacuum contains an non-vanishing number of those particles, per unit of volume. In our example,
they are characterized by the following distribution over the momentum space \cite{Capolupo:2004av}
\be\label{number}
\lfv n(\vk) \rfv = \frac{\sst}{4\pi^3} \frac{\w_1(k)\w_2(k)-m_1 m_2-k^2}{\w_1(k)\w_2(k)}
\ee
with $n(\vk)\equiv\sum_r (a_1^{r\dg}(\vk)a_1^r(\vk)+a_2^{r\dg}(\vk)a_2^r(\vk))$, $k\equiv |\vk|$, and $a_i^{r(\dg)}$
representing ladder operators for particles with well-defined mass.
However, since the physical degrees of freedom of the theory are flavour particles (the only kind of particle that can be produced and detected),
the interpretation as a \textit{gas} or \textit{collection of particles} remains at a mere mathematical level,
the flavour vacuum being absolutely ``empty'' from a physical point of view
(in the sense that no flavoured particles are present in it),
and only characterized by a non vanishing stress-energy tensor expectation value which is detectable via gravitational effects.

The features of the flavour vacuum depend on the model considered and a preliminary investigation on a simple supersymmetric model \cite{Mavromatos:2010ni} showed
that it might behave very differently, according with the spin of the particles involved.

\section{Phenomenology of a SUSY Flavour Vacuum}\label{section:Phenomenology}

\subsection{Free WZ \textit{\`a la} BV}\label{freeWZalaBV}
Our interest in BV formalism was firstly motivated by physics beyond the Standard Model.
The Wess-Zumino model here discussed has been considered in \cite{Mavromatos:2010ni}
after two works in which the \textit{flavour vacuum}
has been regarded as an effective vacuum arising in a string-theoretical framework
\cite{mavrosarkar,Mavromatos:2009rf}. Indeed, a specific model from the \textit{braneworld} scenario, called \textit{D-particle foam model}
\cite{Mavromatos:2006yy,Mavromatos:2005bu,Mavromatos:2008bz,Ellis:2004ay,Ellis:2005ib},
seems to explain neutrino flavour oscillations in terms of flavour oscillation of fundamental strings, in presence of a ``cloud'' (or \textit{foam})
of point-like topological defects in the bulk space. In the spirit of weak coupling string theory, the interaction between the foam
and strings/branes in the theory can be regarded as ``vacuum defects'' from the point of view of a macroscopical observer.
Therefore it has been suggested that BV formalism, together with its ``flavour vacuum'' condensate,
might provide a suitable description of the low energy limit of the model.


In \cite{Mavromatos:2010ni} we presented the behaviour of the \textit{flavour vacuum},
in a simple supersymmetric theory.
The model that was considered involves two free real scalars $S_A(x)$, $S_B(x)$ with mixing,
two free real pseudo-scalars $P_A(x)$, $P_B(x)$ with mixing and
two free Majorana spinors $\psi_A(x)$, $\psi_B(x)$ with mixing:
\begin{multline}\label{lagpsispfg}
\lag =\!\!\!\sum_{\iota=A,B}\Big[
    \partial_\mu S_\iota(x)\partial^{\mu}S_\iota(x)+\partial_\mu P_\iota(x)\partial^{\mu}P_\iota(x)+\\
        +i\bpsi (x)\diracpartial \psi_\iota(x)\Big]-\!\!\!\sum_{\iota,\kappa=A,B}
    \Big[m_{\iota\kappa}^2S^2_\iota(x)+m_{\iota\kappa}^2S^2_\iota(x)+\\
    +m_{\iota\kappa}\bpsi(x)_\iota\psi_\kappa(x)\Big]
\end{multline}
with $m_{AB}=m_{BA}$.
Terms involving products of fields of different flavours disappear when one expresses the model in terms of new
fields, obtained by appropriate rotations of the previous ones:
\bea
\phi_A(x)&=&\ct \phi_1(x)+\st\phi_2(x)\nn
\phi_B(x)&=&-\st \phi_1(x)+\ct\phi_2(x)
\eea
with $\phi=S,P,\psi$, leading to
\begin{multline}\label{lag2}
\lag=
	\sum_{i=1,2}\Big[\partial_\mu S_i(x)\partial^\mu S_i(x)-m_i^2 S_i^2(x)+\\
    +\partial_\mu P_i(x)\partial^\mu P_i(x)-m_i^2 P_i^2(x)+\\
    +\bpsi_i(x)(i\diracpartial-m_i)\psi_i(x) \Big].
\end{multline}
From this latter it is possible to build the usual Fock space for massive particles, previously denoted as
\hilbert{F_0}, which has as ground state the \textit{``massive''} vacuum $\rmv$.
The \textit{flavour} vacuum is hence defined as
\be\label{susyfv}
\rfv\equiv e^{\theta \int d\vx \left(X_{12}(x)-X_{21}(x)\right)} \rmv
\ee
with
\be
X_{12}(x)\equiv \frac{1}{2}\psi^\dagger_1(x)\psi_2(x)+i \dot{S}_2(x) S_1(x)+i \dot{P}_2(x) P_1(x).
\ee
Its features have been explored via its stress-energy tensor expectation value,
being
\begin{multline}\label{set2}
T_{\mu\nu}(x)=\!\!\!\sum_{i=1,2}\!\Big[ 2\partial_{\left(\mu\right.}S_i(x)\partial_{\left.\nu\right)}S_i(x)
    +2\partial_{\left(\mu\right.}P_i(x)\partial_{\left.\nu\right)}P_i(x)+\\
    +i\bar{\psi}_i(x)\gamma_{\left(\mu\right.}\partial_{\left.\nu\right)}\psi_i(x)\Big]-\e_{\mu\nu}\lag.
\end{multline}
It has been shown that the flavour vacuum behaves as a perfect relativistic fluid, \textit{i.e.}
\be
\lfv  T_{\mu\nu}\rfv=diag\{\rho,\pressure,\pressure,\pressure\}
\ee
with
\begin{equation}\begin{split}\label{e1}
\rho\equiv&\lfv T_{00}(x)\rfv=\\
=&\sst \frac{(m_1-m_2)^2}{2 \pi^2}\int_0^K \!\!\!\!dk\;k^2\left(\frac{1}{\w_1(k)}+\frac{1}{\w_2(k)}\right)
\end{split}\end{equation}
\begin{equation}\begin{split}\label{p1}
\pressure\equiv&\lfv T_{jj}(x)\rfv=\\
=&-\sst \frac{m_1^2-m_2^2}{2 \pi^2}\int_0^K \!\!\!\!dk\; k^2 \left(\frac{1}{\w_2(k)}-\frac{1}{\w_1(k)}\right)
\end{split}\end{equation}
$\rho$ representing its energy density, $\pressure$ its pressure and $K$ a momentum cutoff (\textit{cf}
\cite{Mavromatos:2010ni,Capolupo:2010ek}).\footnote{From the perspective of considering
BV formalism as an effective formalism for physics beyond the
Standard Model \cite{mavrosarkar,Mavromatos:2010ni},
such a cutoff must be interpreted as the energy scale up to which the formalism
provides the framework for a good effective theory.}

In particular, disentangling the contribution of the bosonic sector from the fermionic one,
one finds
\begin{multline}\label{energydensityS}
\rho_b=\sst\int_0^K\!\!dk \frac{k^2}{\pi^2}\left(\w_1(k)+\w_2(k)\right)\times\\
\times\left( \frac{\left(\w_1(k)-\w_2(k)\right)^2}{2\w_1(k) \w_2(k)}\right)
\end{multline}
\begin{multline}\label{energydensitypsi}
\rho_f=\sst\int_0^K\!\!dk \frac{k^2}{\pi^2}\left(\w_1(k)+\w_2(k)\right)\times\\
\times\left( \frac{\w_1(k)\w_2(k)-m_1 m_2-k^2}{\w_1(k)\w_2(k)}\right)
\end{multline}
\be
\pressure_b=-\rho_b,\;\;\;\;\;\;\pressure_f=0
\ee
in which the standard normal order has being adopted.

\subsection{Flavour vacuum as a source of Dark Matter}

An important result emerges from the above outlined analysis: the equation of states $w\equiv\pressure/\rho$
for the bosonic and the fermionic sectors are different, $w_b=-1$ and $w_f=0$ holding.
The emphasis on the novelty of this SUSY breaking mechanism has already being remarked \cite{Mavromatos:2010ni}.
We are now aimed to explore the interesting phenomenology connected with such a result.
Our simple model implies a physical vacuum that is a combination of two fluids which behave quite differently:
both permeate the empty space uniformly and statically, but one has a cosmological-constant-like behaviour ($w=-1$),
while the other behaves as dust ($w=0$).
The role of the flavour vacuum as source of Dark Energy (which now is played only by the bosonic sector of the theory)
has been extensively discussed in literature
\cite{Blasone:2004hr,Blasone:2004yh,Capolupo:2006et,Capolupo:2006re,Capolupo:2008rz,Capolupo:2007hy,Blasone:2007iq,Blasone:2008rx,Blasone:2007jm,Blasone:2006ch}.
Here we present a new feature of the flavour vacuum:
its contribution to Dark Matter.

Dark Matter is the name given to unknown sources of gravitational effects, whose presence, primarily within and
around galaxies, has been established by many astrophysical data \cite{Bertone:2004pz,Jungman:1995df}.
Numerical simulations of structure formation have shown that ``hot'' (relativistic)
particles cannot explain the observed structures at galactic scales,
therefore Dark Matter is expected to be made out of fairly massive and ``cold''
(non-relativistic) particles. Big-Bang nucleosynthesis limits on the average
baryonic content of the Universe exclude that (the majority of) Dark Matter
is made out of ordinary baryonic matter (\textit{i.e.} atoms). Furthermore, although
``dark'', in the sense that does not emit nor absorb light (\textit{i.e.} electromagnetically
neutral), Dark Matter might couple to ordinary matter in other ways
(besides gravity); however, arguments on its density and thermal production
at early times imply that such a coupling must be weak.

Both astrophysics and particle physics have been proposing suitable candidates
for Dark Matter through the last three decades, giving rise to an enormous
wealth of choice. However, because of the absence of direct detections and the lack of predictions
by theoretical models, plagued by an undesirable abundance of free parameters,
the nature of Dark Matter remains elusive.

The fermionic sector of the flavour vacuum in the model here presented clearly fulfills basic requests for
a Dark Matter candidate:
it contributes to the energy content of the universe; it is ``dark'' (\textit{i.e.} it is an electromagnetically neutral object,
since (s)neutrino fields do not couple with the electromagnetic field);
furthermore, it does not interact with any other of the SM particles (excluding gravitational effects), being the 
\textit{empty} state for the (s)neutrino sector; unlike its bosonic counterpart, it is purely \textit{pressureless}.\footnote{This
is certainly true for the free above mentioned model; the
possibility of extending this result to interactive models will be discussed later on.}

A possible concern about its uniform distribution in space, in contrast with the observed distribution of Dark Matter which
is usually gathered in clusters around and inside galaxies, can be easily dispelled 
by recalling that we are actually modeling a simple ``empty'' universe.
If a non-uniform matter distribution is considered in our toy universe in addition to the flavour vacuum, it would start to
interact gravitationally with our vacuum condensate. Thanks to initial irregularities in the matter distribution, we expect them to form clusters
via \textit{gravitational instability} (gravity tends to enhance irregularities, pulling
matter towards denser regions \cite{Liddle:2009zz}), as the system evolves with time.
It is known that such an effect, on the other hand, does not necessarily occurs for Dark Energy-like fluids
\cite{amendola}, as the bosonic component of the flavour vacuum,
which can persist in their state of spatial uniformity even in presence of clustered matter.
The evolution of our flavour vacuum, considering both its bosonic and fermionic components, in presence
of other matter and gravitational interaction, represents necessarily an object of future studies.

\subsection{Testability}

An interesting aspect of the model concerns the interplay between the two fluids.
Supersymmetry imposes that the energy density of the bosonic component is tied up
to the energy density of the fermionic component; in a more realistic theory, therefore,
one should be able to reproduce the current experimental value of the ratio between the Dark Energy density and
the Dark Matter energy density ($\sim 2.8$), in the optimist belief that the flavour vacuum is the only responsible for
both of them. The role of a curved background in the formulation of the theory might be crucial,
since the energy density of a dust-like fluid gets diluted by the expansion of the universe, whereas such an effect
does not occur for a cosmological-constant type,
and therefore the ratio between those two quantities changes dramatically with time.

Within a momentum cutoff regularization framework, as the one here presented,
the two energy densities depend on such a cutoff, which is the \textit{same} for both quantities.
The ratio between them can in general be cutoff dependent, as it actually is in the case
here presented. On one side, one might hope that in a more realistic theory (on a curved background, for instance)
the ratio might be cutoff independent.
On the other hand, one could consider the opposite situation, in which the ratio varies with the cutoff, as highly desirable:
if the ratio is fixed from the cutoff, the same value for the cutoff would also fix the value of the energy.
This implies that once the cutoff is decided on the basis of experimental data on the ratio,
the model gives a precise prediction for the absolute values of the energy densities,
which can be compared with their observational estimates.

In order to illustrate these ideas we will present a concrete example.
Let us assume that our supersymmetric model is effective up to the energy scale $K$ (which comes from deeper theories,
as, for instance, in \cite{Mavromatos:2010ni} ).
In the standard Big-Bang  picture, this means that when the universe cools down to that energy,
the flavour vacuum starts to be the effective description of the vacuum state of the (unknown) underlying theory.
We call $t_0$ the time corresponding with this transition and $a_0$ the corresponding scale factor.

In our toy universe, we assume that  at $t_0$, in absence of any other sources of energy or matter,
the energy/matter content of our toy universe is only due to the flavour vacuum. Moreover, we assume that it can be describe,
at a classic level (\textit{i.e.} on sufficiently large scales),  in terms of two fluids:
a first one, due to the bosonic component of the flavour vacuum and described by $\rho_b$ and $w=-1$,
and a second one, due to its fermionic component and described by
$\rho_f$ and $w=0$.
We will regard the bosonic component as the only source of Dark Energy and the fermionic as the only source of Dark Matter.
Both $\rho_b$ and $\rho_f$ are function of the following parameters:
(s)neutrino masses, mixing angles, and the cutoff $K$.
If we know (from observations) the neutrino masses and mixing angles, and we can constrain parameters induced
 by SUSY breaking, the cutoff is the only parameter left to determine.

As our toy universe expands, we assume that the two fluids obey Einstein equations and therefore they
scale as
\be\begin{split}\label{dedmev}
\rho_f(t)\;\;a(t)^3=&\rho_f \;\;a_0^3\\
\rho_b (t)=&\rho_b.\\
\end{split}\ee
This means that \textit{today} their value is
\be\begin{split}\label{dedmnow}
\rho_f(t_{now})=&\rho_f \;\;a_0^3\\
\rho_b(t_{now})=&\rho_b.\\
\end{split}\ee
respectively, being $a(t_{now})=1$ by convention.
Those two quantities depends on the following parameters: (s)neutrino masses, mixing angles, cutoff energy, scale
factor at $t_0$.
Provided with these expressions, we can then test our model in two ways.
\begin{enumerate}
	\item If observational data enable us to constrain (s)neutrino masses, mixing angles, Dark Matter and Dark Energy densities,
	from (\ref{dedmnow}) we can derive the other parameters left: the cutoff energy and the scale factor $a_0$.
	Well equipped with all the parameters of the theory, we will then be able to
	check if the model is in reasonable agreement with other standard models.
	For example, if the scale factor $a_0$ fitting all data corresponds to a time \textit{in the future} (for $a_0>1$),
	the model has to be rejected, or corrected at least.
	\item On the other hand, theoretical reasons might suggest specific values for the cutoff (if for instance the flavour
	vacuum rises in the low energy limit of an underlying theory, and/or the scale factor $a_0$, being the temperature of the universe
    inversely proportional to its scale factor.
	In this case, we might be able to make \textit{a prediction} on the value of the Dark Matter and Dark Energy density,
	via formulae (\ref{dedmnow}), that might be compared with observational estimates. On this basis
	our model is therefore accepted or refused.
\end{enumerate}

\subsection{A Preliminary Test}

The simple toy model discussed in Section \ref{freeWZalaBV} is not realistic enough to hold the comparison
with data already available: only two generations of neutrinos have been considered,
neither matter or interactions are present,
SUSY is unbroken, there is no prescription for the cutoff $K$.\footnote{The
assumption of treating neutrinos as Majorana particles might also be questioned.}
However, some preliminary tests can be performed.

As just explained, in a realistic theory with three generations instead of two, it is possible to
constrain the parameter space of mixing angles and masses thanks to observational data.
In absence of such a theory, we will limit our selves to check if our simpler model
admits a choice of parameters that gives rise to physically ``plausible'' estimations for Dark Energy and
Dark Matter densities.
More precisely, we shall check the compatibility of our model with the relation
\be\label{energyscaleobs}
\rho_\Lambda \sim \rho_{DM} \sim (\Delta m_{ij}^2)^2,
\ee
with $\rho_{\Lambda/DM}$ the Dark Energy/Matter density today
and $\Delta m_{ij}$ the difference of the squared masses of
neutrinos\footnote{The energy scale of Dark Energy is far away from all natural scales provided by
the SM via particle masses. Only one fundamental scale is known to be comparable with the Dark
Energy one: the scale of neutrino physics.
Boundary on total masses of neutrinos show that they are much
lighter then all other particles:	astrophysical data indicate that $\Sigma\;m_\nu<058\;eV$, with $95\%$ of confidence \cite{Komatsu:2010fb}
(the sum runs over all possible species - possibly more then three - that where present in the early Universe).
Moreover, direct observations on solar and atmospheric neutrinos show that
$\Delta{m^2_{12}}\approx8\cdot10^{-5}\;eV^2$ and $\Delta{m^2_{23}}\approx2.5\cdot10^{-3}\;eV^2$ (being
$m_i^2-m_j^2\equiv\Delta m_{ij}^2$, \textit{cf} \cite{Bilenky:2010zza} and  references therein).
These mass scales $10^{-1}\div10^{-2}eV$ have to be compared with the scale $10^{-3}eV$,
that one obtains from $\rho_{\Lambda}=3\times10^{-11}eV^4$.
This ``coincidence'' gave rise to many works, besides the ones connected with BV formalism,
aimed to provide a theoretical explanation for it (see \cite{Gurwich:2010gb} and references therein).\\
The more famous \textit{coincidence problem} regarding Dark Energy
concerns the similar density of Dark Energy and Dark Matter ($\rho_{\Lambda}\approx 2.8\rho_{DM}$) as measured today,
which requires a notable fine-tuning of initial conditions considering their very different evolution in time.\\
These two ``coincidences'' are combined together in formula (\ref{energyscaleobs}).},
which for our model becomes
\be\label{energyscale}
\rho_b \sim \rho_f(t_{now}) \sim (\Delta m^2)^2,
\ee
in the assumption that all Dark Energy and Dark Matter of our toy universe is due to the flavour
vacuum. 
In other words, is there a sensible region of the parameter
space that gives rise to (\ref{energyscale})?

Once provided with a realistic theory, the reasoning goes the other way around: given the space of
parameters constrained by observational data, does (\ref{energyscale}) hold?
However, the analysis on which we are embarking is neither irrelevant nor negligible:
previous analyses hardly conciliate
the very different scales of energies entering into the problem, such as the momentum cutoff,
which presumably is greater then the $TeV$ scale, and
neutrino mass differences (\textit{cf} for instance \cite{Capolupo:2008rz}).
The aim of this section is therefore to show that even in our simple toy model, non-perturbative formulae
describing the features of the flavour vacuum can accommodate very different scales in a natural way, giving
rise to physically sensible values for Dark Energy and Dark Matter densities.

Recapitulating, in the following we shall assume that
the physical vacuum is effectively described by
the flavour vacuum defined by (\ref{susyfv}) for energies lower than $K$; over large distance scales,
such a flavour vacuum behaves as a classical fluids,
obeying Einstein equations (\textit{i.e.} (\ref{dedmev}) and (\ref{dedmnow}) hold).
We will further assume that some radiation and matter are present in our toy universe, whose density
is at least one order lower than the flavour vacuum density. Their presence
justifies the notion of ``temperature'' and defines the profile of the time-evolution of our
toy universe. Since we assume the neutrino sector not being coupled with any other fields,
the flavour vacuum and the matter/radiation content of the universe interact
only gravitationally. As mentioned, we expect this interaction to lead the fermionic flavour vacuum to cluster
together with ordinary matter, leaving the bosonic flavour vacuum homogeneously distributed.
These effects are reasonably expected
as long as gravitational effects are relevant only on cosmological scales, at which the flavour vacuum
is well approximated by a classical fluid.
As a consequence of a possible non-uniform distribution in space of the fermionic fluid,
we shall consider the value of (\ref{energydensitypsi}) not as a local attribute but as a global,
or ``averaged'' over sufficiently large scales, property.

In order to check the compatibility of our model with (\ref{energyscale}),
we start by defining the quantity $\xi\equiv (1-m_2^2/m_1^2)^2$, with $m_2$ the smaller
of the two masses ($m_2<m_1$). Since $0<\xi<1$, we can expand (\ref{energydensityS}) and (\ref{energydensitypsi}) in series around $\xi=0$ and get to
\bea\label{rhobrhof}
\rho_b &=& \frac{\sst}{\pi^2}m_1^4 f(K/m_1) \xi+\mathcal{O}(\xi^{3/2})\nn
                &\approx& \frac{\sst}{\pi^2}(\Delta m^2)^2 f(K/m_1)\nn
\rho_f &=& \frac{\sst}{\pi^2}m_1^4 g(K/m_1) \xi+\mathcal{O}(\xi^{3/2})\nn
                &\approx& \frac{\sst}{\pi^2}(\Delta m^2)^2 g(K/m_1)
\eea
with
\bea\label{f&g}
f(K/m_1)&=&\int_0^{K/m_1}dx\frac{x^2}{4(1+x^2)^{3/2}}\\
g(K/m_1)&=&\int_0^{K/m_1}dx\frac{x^4}{4(1+x^2)^{3/2}}
\eea
and $m_1^4 \xi=(\Delta m^2)^2\equiv(m_1^2-m_2^2)^2$.
Relations (\ref{rhobrhof}) are good approximations of the exact values,
as long as the two masses are very similar $m_1\sim m_2$.
All divergencies connected with our problem are included in function $f(K/m_1)$ and $g(K/m_1)$, for their argument
running to infinity.
In the following analysis a physical cutoff of momenta ($K$, rescaled by the neutrino mass $m_1$)
will be considered, in the belief that flavour physics \textit{\`a la} BV must be regarded
as an effective description at low energy scales of a deeper theory \cite{mavrosarkar}.
Clearly other renormalization tools might be required if this assumption is dropped,
for instance, in a pure self consistent quantum field theoretical approach.
Despite this choice, it is remarkable, however, that the relation $\rho_{b/f}\propto (\Delta m^2)^2$ has been derived
entirely analytically.

In the following, we would like to show that in a cutoff-regularization scheme the two functions in (\ref{f&g})
can give rise to physically sensible values, \textit{i. e.} the function $f(K/m_1)$ remains relatively small
even when the cutoff is very high (giving rise to the hierarchy: high cutoff/low Dark Energy density),
whereas $g(K/m_1)$ can be considerably greater then $f(K/m_1)$ for the same choice of cutoff,
motivating the observed discrepancy between the Dark Energy and Dark Matter densities
at early times.
We should stress once more that a strict comparison with available experimental data would be possible
once a more realistic model will be given.

\begin{figure}
\includegraphics[width=2.25in]{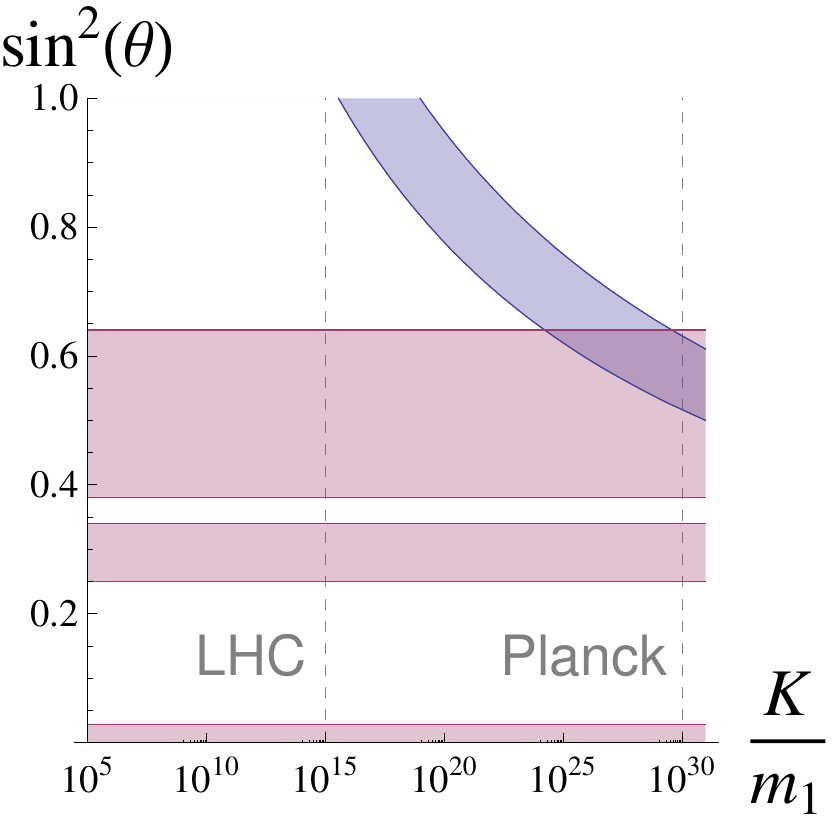}\caption{\label{figure1}The parameter space $(K/m_1,\sst)$
is plotted. In blue, the acceptance region for the condition (\ref{DEcondition}) (boundaries of the region
correspond to $lhs=0.9$ and $lhs=1.1$). Physical mixing angles for
three generation of neutrinos are: $\sin^2 \theta_{13}= 0.000^{+0.028}$, $\sin^2 \theta_{12}=0.30^{+0.04}_{-0.05}$,
$\sin^2 \theta_{23}=0.50^{+1.4}_{-1.2}$ (within $2 \sigma$) \cite{Giunti:2007ry}.
These values are also plotted in the graph (red regions). Dotted lines represents LHC energy scale ($\sim 1TeV$) and the Planck energy scale ($\sim10^{28}eV$)
fixing $m_1=10^{-2}eV$.}
\end{figure}

We will now focus on the former of (\ref{rhobrhof}). Being SUSY unbroken in our toy model,
neutrinos and sneutrinos have the same masses.
So the $m_1$ and $m_2$ appearing in (\ref{rhobrhof}) are the masses of two neutrinos, even though $\rho_b$
encodes the contribution of the bosonic sector of the theory, which in a realistic case would be affected
by the breaking of SUSY via effective masses greater than $m_i$.
Recalling the observed relation (\ref{energyscaleobs}), we wonder now if is there any region of the parameter space ($\theta$,$K$) that might generate a similar situation (\textit{i.e.} $\rho_b\sim (\Delta m^2)^2$) in our model.
Quite interestingly, the condition
\be\label{DEcondition}
\frac{\sst}{\pi^2}f(K/m_1)\sim 1
\ee
is satisfied by a region of the plane ($\theta$,$K$), which is not that far from the expected value for
a realistic theory. As shown in Figure \ref{figure1}, if the cutoff of the theory lies somewhere
between the TeV ($10^{12}\;eV$) scale and the Plank scale ($10^{28}\;eV$), the value of $\sst$ must
be around the $0.5 \div 1$ region. Moreover, a complete overlap with one real mixing angle is obtained
in a region with a very high cutoff, close to the Planck scale.
It should be emphasized that the existence itself of such a region is highly non-trivial, since it appears from
the combination of parameters spanning a wide range of energies, being $\Delta m^2\sim 10^{-4}eV^2$
and $K>10^{12}\;eV$.

Focusing now on the latter of (\ref{rhobrhof}), we shall proceed with a similar analysis in order to test the hypothesis
that the fermionic sector of the model provides a sensible Dark Matter candidate.
As explained in the previous section, under the assumption that the flavour vacuum behaves
as a perfect classical fluid on large scales, the fermionic contribution would get diluted
with time as the universe expands. Its value \textit{today} would then be
\be\label{rhofnow}
\rho_f(t_{now})=\rho_f \;\;a_0^3
\ee
with $a_0$ the scale factor corresponding to the time at which the model became effective.
In order to reproduce the relation $(\Delta m^2)\sim \rho_f(t_{now})$
we expect
\be
 a_0^3 \frac{\sst}{\pi^2} g(K/m_1)\sim 1
\ee
that can be obtained by combining (\ref{rhofnow}), (\ref{energyscale}) and (\ref{rhobrhof}).
Because of the constrain on the Dark Energy density
(\ref{DEcondition}) the above condition becomes
\be\label{DMcondition}
 a_0^3 \frac{g(K/m_1)}{f(K/m_1)}\sim 1.
\ee
We already know that $a_0$ must be extremely small, but what is the ratio between
$g(K/m_1)$ and $f(K/m_1)$ for large $K/m_1$ ($K/m_1>10^{14}$)?
\begin{figure}
\includegraphics[width=2.25in]{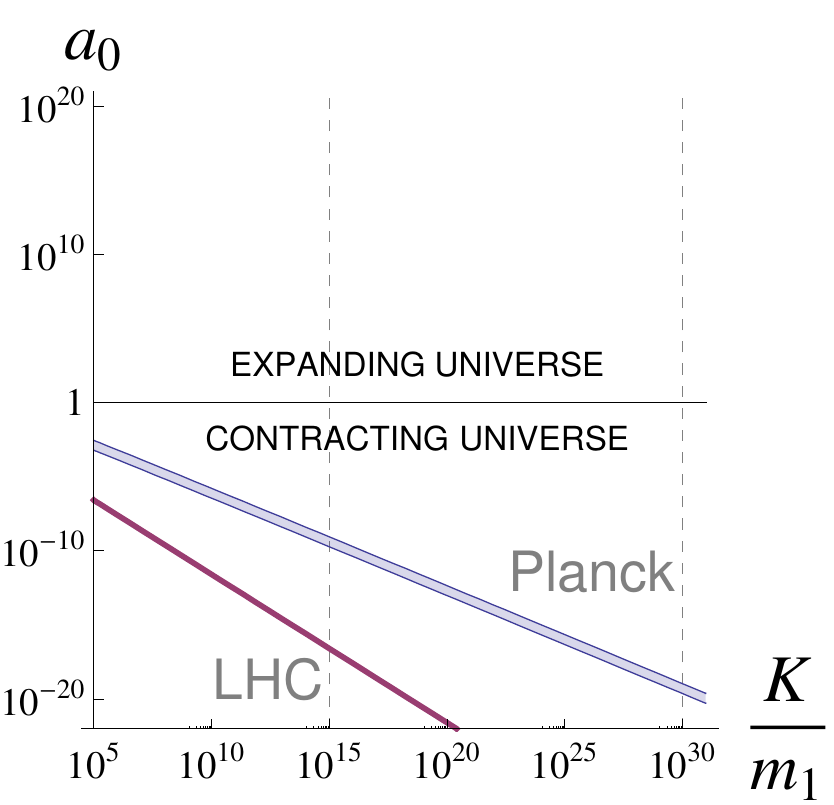}\caption{\label{figure2}
The parameter space $(K/m_1,a_0)$ is plotted.
In blue, the acceptance region for the condition (\ref{DEcondition}) (boundaries of the region
correspond to $lhs=10$ and $lhs=0.1$).
The red line depicts the scale factor of the real universe as a function of its average temperature,
fixing $m_1=10^{-2}eV$. In a more realistic theory, the acceptance region (which may differ from the one
here presented) is further reduced by comparison of the first of (\ref{rhobrhof}) with Dark Energy data,
which gives a constraint on possible momentum cutoffs;
the model is ruled out if the resulting region lies \textit{below}
the red line, since it would predict an amount of Dark Matter greater than what is observed;
the model is consistent with data on Dark Matter if the region overlaps or lies \textit{above}
the red line; in this latter case the flavour vacuum contributes only to a \textit{fraction}
of total Dark Matter density.}
\end{figure}
As shown in Figure \ref{figure2}, equation (\ref{DMcondition}) is indeed satisfied for large values of $K$
and very small values of $a_0$, as one would expect from a realistic theory. If, for instance, the cutoff is
set equal to the Planck scale, relation (\ref{DMcondition}) is satisfied for $a_0=10^{-20}$,
which sets the transition phase (when the flavour vacuum became effective)
well far in the past, when our toy universe was $10^{20}$ times smaller than ``now''.
Once more, parameters characterized by very different values ($a_0\sim 10^{-20}$, $K/m_1\sim 10^{30}$)
combine together to give rise to a third scale, which has a physical significance (nowadays Dark Matter density).

Despite the unrealistic nature of our toy model, these first quantitative tests go in the
right direction and certainly motivate the study of more realistic models, which hopefully will share
with our toy model all the good features here discussed.

\section{Towards Interactive Flavour Vacua}\label{section:Interactions}

\subsection{A new method of calculation: Free WZ Revised}\label{ssANMOC}
Following the standard literature, the results in \cite{Mavromatos:2010ni} discussed so far have been derived by the following approach:
\begin{enumerate}
 \item stress-energy tensor has been written in terms of massive fields;
 \item massive fields have been decomposed in terms of massive ladder operators;
 \item massive ladder operators have been written in terms of flavour ladder operators;
 \item hence, stress-energy tensor has been written in terms of flavour ladder operators;
 \item flavour-vev of the stress-energy tensor has been reduced by acting with the flavour ladder operators on the flavour vacuum.
\end{enumerate}
However, such a long procedure can be avoided and the same exact result may be obtained in a much shorter way, following the steps:
\begin{enumerate}
 \item flavour vacuum is written as $\rfv=G^{\dagger}_\theta \rmv$, the flavour-vev of $T_{\mu\nu}$ becoming a vev of $\mathcal{O}\equiv G_{\theta} T_{\mu\nu} G^\dagger_{\theta}$;
 \item the stress-energy operator is transformed under the action of the $G_{\theta}\slot G^\dagger_{\theta}$, using (\ref{G1WT}); the transformed operator $\mathcal{O}$ will be
  expressed in terms of the flavour fields, rather then the massive ones;
 \item (\ref{flavourandmassive}) is used to write the flavour fields in terms of the massive fields; $\mathcal{O}$ is again expressed in terms of the massive fields, but this time
  the operator $G_{\theta}$, and its complicated exponential structure, is not present any more;
 \item  a vev of $\mathcal{O}$, expressed as a simple combination of massive fields, is left, which can be reduced by decomposing the massive fields into massive ladder operators.
\end{enumerate}

As a neat example of the above procedure, we will derive the results of Section \ref{freeWZalaBV}
via this new method.
Recalling the discussion in Section \ref{freeWZalaBV},
in the study of the free WZ model we can consider the bosonic and the fermionic component separately,
by evaluating relevant quantities in two separated contexts (a bosonic theory and a fermionic one) and eventually combining together the results.
Furthermore, the pseudoscalar and the scalar field are indistinguishable for our purposes, therefore
we are allowed to consider just the scalar field, keeping in mind to sum its contribution to the relevant quantities twice.

In the real scalar case, we have
\be
T^b_{00}(x)=\sum_i\left(\pi_i^2(x)+\left(\vec{\nabla}\phi_i(x)\right)^2+m_i^2\phi_i^2(x)\right)
\ee
with $\pi_i\equiv\dot{\phi}_i$, the conjugate momentum of $\phi_i$.
Since
\be
G_{\theta}(t)=e^{i\theta\int d\vx \left( \pi_2(x)\phi_1(x)-\pi_1(x)\phi_2(x) \right)}
\ee
from which
\begin{equation}\begin{split}
G_{\theta}(t) \phi_1(x)  G^{\dagger}_{\theta}(t)=\phi_1(x)   \ct -\phi_2(x)  \st\\
G_{\theta}(t)\phi_2(x)  G^{\dagger}_{\theta}(t)=\phi_1(x)   \st +\phi_2(x)  \ct
\end{split}\end{equation}
and
\begin{equation}\begin{split}
G_{\theta}(t) \pi_1(x)  G^{\dagger}_{\theta}(t)=\pi_1(x)   \ct -\pi_2(x)  \st\\
G_{\theta}(t)\pi_2(x)  G^{\dagger}_{\theta}(t)=\pi_1(x)   \st +\pi_2(x)  \ct
\end{split}\end{equation}
via the Baker-Campbell-Hausdorff formula
\be\label{BCH}
e^Y X e^{-Y}=X+[Y,X]+\frac{1}{2}[Y,[Y,X]]+\frac{1}{3!}[Y,[Y,[Y,X]]]+...
\ee
we can write
\be
 G_{\theta}(t)   \left(\sum_{i=1,2}\pi_i^2(x)\right)   G_{\theta}^\dagger (t)= \left(\sum_{i=1,2}\pi_i^2(x)\right),
\ee
\begin{multline}
G_{\theta}(t)  \left(\sum_{i=1,2}\left(\vec{\nabla}\phi_i(x)\right)^2\right)   G_{\theta}^\dagger (t) =\\
\left(\sum_{i=1,2}\left(\vec{\nabla}\phi_i(x)\right)^2\right)
\end{multline}
and
\begin{multline}
\lmv G_{\theta}(t) (m_1^2\phi_1^2(x)+m_2^2\phi_2^2(x)) G_{\theta}^\dagger (t) \rmv=\\
								\lmv  (m_1^2\phi_1^2(x)+m_2^2\phi_2^2(x))  \rmv+\\
													\sst(m_1^2-m_2^2)\lmv(\phi_2^2(x)-\phi_1^2(x))\rmv.
\end{multline}
It follows that
\begin{multline}
\lfv T_{00}(x) \rfv=\lmv T_{00}(x)\rmv +\\
		  +\sst(m_1^2-m_2^2)\lmv(\phi_2^2(x)-\phi_1^2(x))\rmv
\end{multline}
and therefore
\begin{multline}\label{energybospartiv}
\rho_b=\lfv: T_{00}(x) :\rfv=\\
      =\sst(m_1^2-m_2^2)\lmv(\phi_2^2(x)-\phi_1^2(x))\rmv
\end{multline}

Equivalently for $T_{jj}(x)$ we have
\begin{multline}\label{pressurebospartiv}
\pressure_b=\lfv: T_{jj}(x) :\rfv=-\sst(m_1^2-m_2^2)\times\\
					\times \lmv(\phi_2^2(x)-\phi_1^2(x))\rmv=-\rho_b.
\end{multline}
Once the fields in (\ref{energybospartiv}) and  (\ref{pressurebospartiv}) are decomposed in terms of the ladder operators and
the quantum algebra is simplified, expressions (\ref{energydensityS}) and (\ref{energydensitypsi}) are correctly reproduced.

In the fermionic case, a similar procedure leads to
\begin{multline}\label{energyferpartiv}
\rho_f=\lfv: T_{00}^{f}(x):\rfv=\sst\left(m_1-m_2\right)\times\\
				\times \lmv\left( \bpsi_2(x) \psi_2(x)-\bpsi_1(x) \psi_1(x)\right)\rmv
\end{multline}
and
\be\label{pressureferpartiv}
\pressure_f=\lfv: T_{jj}^{f}(x):\rfv =0.
\ee
By comparing (\ref{energyferpartiv}) and (\ref{energybospartiv}), the analogy between the fermionic and the bosonic condensate that earlier
was hidden in formulae (\ref{energydensityS}) and (\ref{energydensitypsi}) is now more evident.
Again, formula (\ref{energydensitypsi}) is correctly reproduced, once
the operatorial structure of the fields is simplified with respect to $\lmv \slot \rmv$.
The expression (\ref{energyferpartiv}) dispels any doubts concerning formula (\ref{energydensitypsi}) and its possible dependency on the
specific form of the gamma matrices and spinors used to achieve the results of \cite{Mavromatos:2010ni},
being (\ref{energyferpartiv}) independent of such a choice \cite{Peskin:1995ev}.

Furthermore, Supersymmetry enables us to rewrite this result in terms of the bosonic fields only.
For the \textit{massive} vacuum $\rmv$ we know that
\be
\lmv T_{\mu\nu}(x)\rmv=0
\ee
which leads to
\be
\lmv \bpsi_i(x) \psi_i(x) \rmv=-4m_i\lmv \phi^2_i(x)\rmv
\ee
and hence
\be
\rho_{WZ}=2\sst(m_1-m_2)^2\lmv \left(\phi_1^2(x)+\phi_2^2(x)\right)\rmv
\ee
\be
\pressure_{WZ}=-2\sst(m_1^2-m_2^2)\lmv(\phi_2^2(x)-\phi_1^2(x))\rmv
\ee
in accordance with (\ref{e1}) and (\ref{p1}).

The procedure exemplified in the previous section can be easily implemented
for other fields: one might want to consider Dirac or two component
Weyl spinors as well as complex scalar fields, getting to analogous results.
For mere speculative reasons, applications to vector fields or even more complex
objects might be thought: the method involves a manipulation of the
stress-energy tensor, with the use of equation of motion of the field and its
(anti-)commutation rules, regardless of the tensorial or spinorial structure of
the field itself.
Furthermore, extending these results to more then two flavours is rather straightforward.

Finally, the method enables us to distinguish among all the terms
of the stress-energy tensors the ones that really contribute to the final result.
This might be helpful in understanding the behaviour of the flavour vacuum in
more realistic theories, as we shall see in the forthcoming sections.\footnote{In addition, we would like to stress that
the method does not require an explicit decomposition of the flavour fields in terms
of flavour ladder operators. Such a decomposition has been object of a debate in literature,
raised by the authors of \cite{Fujii:1998xa}.
Although the problem was exhaustively discussed in \cite{Blasone:1999jb}
and \cite{Fujii:2001zv}, not all the community was convinced by the arguments presented \cite{Giunti:2003dg}.
Without entering into the details of the dispute, here we would like to suggest that a different point of view on the formalism,
such as the one offered by formulae (\ref{energyferpartiv}) and (\ref{energybospartiv}),
where an observable quantity concerning a flavour state has been calculated
without the explicit use of the controversial decomposition, might help in a deeper understanding of the problem
and the formalism itself.}

\subsection{Self-interactive Bosons}\label{ssSIB}

An example of the applications just discussed is offered by a $\lambda \phi^4$ model.
The theory
\be\label{phiforthlinear}
\lag =\sum_{i=1,2}\left(\partial_\mu \phi_i \partial^\mu \phi_i-m_i^2 \phi_i^2-\lambda \phi_i^4\right)
\ee
can be regarded as derived from a model with flavour mixing:
\begin{multline}\label{flavourphiforth}
\lag = \partial_\mu \phi_A \partial^\mu \phi_A+\partial_\mu \phi_B \partial^\mu \phi_B-m_A^2 \phi_A^2-m_B^2 \phi_B^2+\\
    -m_{AB}^2\phi_A\phi_B
			-\!\!\!\!\!\!\!\!\sum_{\iota,\kappa,\lambda,\rho=A,B}\!\!\!\!\!\!\!\!
									g_{\iota\kappa\lambda\rho} \phi_\iota \phi_\kappa \phi_\lambda \phi_\rho
\end{multline}
with the usual rotation
\be\label{usualrotation}\begin{split}
\phi_A=\ct \phi_1-\st \phi_2\\
\phi_B=\st \phi_1+\ct \phi_2
\end{split}\ee
and a specific choice of the coupling constants $g_{\slot}$.
Since the expression of $G_{\theta}$ in terms of the fields can be deduced from
\be\begin{split}
G_{\theta}^{\dagger}\phi_1G_{\theta}=\ct \phi_1-\st \phi_2\\
G_{\theta}^{\dagger}\phi_2G_{\theta}=\st \phi_1+\ct \phi_2
\end{split}\ee
just using commutation relations between fields and conjugate momenta,
which are not modified by the form of the Lagrangian \cite{Peskin:1995ev},
expression
\be
G_{\theta}=e^{i\theta\int d\vx (\dot{\phi}_2 \phi_1-\dot{\phi}_1 \phi_2)}
\ee
that was found valid in the free case, holds also in the interactive one.

If we \textit{assume} that the flavour vacuum is defined as
\be
\rfv\equiv G^{\dagger}_{\theta}\rmv
\ee
na\"ively generalizing the free case, with $\rmv$ the ground state of the theory described by (\ref{phiforthlinear}),
we can easily see that
\begin{multline}
\lfv T_{\mu\nu} \rfv=\lmv T_{\mu\nu} \rmv+\\
    +\e_{\mu\nu}\left(\sst (m_1^2-m_2^2)\lmv \phi_2^2-\phi_1^2\rmv\right)+\\
	+\e_{\mu\nu}\lambda \lmv\left( (\phi_2 \cos (\theta)-\phi_1 \sin (\theta ))^4+\right.\\
				+\left.(\phi_1 \cos (\theta)+\phi_2 \sin (\theta ))^4 -\phi_1^4-\phi_2^4 \right)\rmv.
\end{multline}
We can therefore state that the equation of state is given by
\begin{multline}\label{wphiforth}
w=\frac{\lfv : T_{jj}:\rfv}{\lfv : T_{00}:\rfv}=\\
  =\frac{-\fvev{\sum_{i=1,2}(m_i^2\phi_i^2+\lambda\phi_i^4)}}{\fvev{\sum_{i=1,2}(m_i^2\phi_i^2+\lambda\phi_i^4)}}=-1
\end{multline}
in which
\begin{multline}\label{fvevdef}
\fnormal{f(\varphi_1,\varphi_2)}\equiv \\
f(\ct \varphi_1-\st \varphi_2,\st\varphi_1+\ct\varphi_2)- f(\varphi_1,\varphi_2).
\end{multline}

Quite notably, this result generalizes the analogous result for the free theory,
in a completely \textit{non-perturbative} way:
equation (\ref{wphiforth}) is independent of the explicit form of the fields and the ground state $\rmv$, which we might be able to recover
just in a perturbative treatment of the model.

In fact, it is possible the further generalize the above result for \textit{any} interactive theory for two scalar fields with flavour mixing
in the following form:
\begin{multline}
\lag = \partial_\mu \phi_A \partial^\mu \phi_A+\partial_\mu \phi_B \partial^\mu \phi_B-m_A^2 \phi_A^2-m_B^2 \phi_B^2+\\
    -m_{AB}^2\phi_A\phi_B+\lag_{int}(\phi_A,\phi_B)
\end{multline}
with $\lag(\phi_A,\phi_B)$ any polynomial function of $\phi_A$ and $\phi_B$. It is easy to show that
\be
\lfv :T_{\mu\nu}: \rfv=\e_{\mu\nu}\lmv \fnormal{\sum_{i=1,2}m_i^2 \phi^2_i-\lag_{int}}\rmv
\ee
leading \textit{always} to the equation of state $w=-1$.

\subsection{Self-interactive fermions}

Analogously, we can generalize the result presented in Section \ref{freeWZalaBV} for fermionic fields (namely, $w=0$)
for a certain class of self-interactive theories.
We start by considering a theory written in terms of the massive fields $\psi_1$ and $\psi_2$:
\be\label{lag}
\lag=\sum_i \bpsi_i (i\diracpartial-m_i)\psi_i+\lag_{int}
\ee
with $\lag_{int}$ a suitable polynomial function of $\psi_i$ and $\bpsi_i$.
Again, we regard (\ref{lag}) as the \textit{diagonalized} Lagrangian:
in case of flavour mixing, $\psi_1$ and $\psi_2$ come from a rotation of the flavoured fields $\psi_A$ and $\psi_B$.

Combining our previous discussion on the bosonic case and results of Section \ref{ssANMOC},
we can write
\bea
\lfv : T_{00} : \rfv&=& \fvev{T_{00}} =\nn
		    &=&\fvev{\sum_i m_i \bpsi_i \psi_i+\lag_{int}}
\eea
in which we used
\be	
\fnormal{\bpsi_i \vec{\gamma}\cdot\vec{\partial} \psi_i}=0.
\ee

Analogously, the $jj\neq 00$ component of the stress energy tensor is given by
\be
T_{jj}=\sum_i\left(\bpsi_i\gamma_j \partial_j\psi_i\right)-\e_{jj}\lag
\ee
that on-shell can be written as
\be
T_{jj}=\sum_i\left(\bpsi_i\gamma_j \partial_j\psi_i+\psi^\dagger_i\left[L_{int},\psi_i\right]\right)+\lag_{int}
\ee
with $L_{int}=\int\; d^3x \lag_{int}$,
leading to
\begin{multline}
\lfv : T_{jj} : \rfv= \fvev{T_{jj}} =\\
=\fvev{\sum_i \bpsi_i \left[L_{int},\psi_i \right]+\lag_{int}}
\end{multline}

We can now distinguish two cases:
\begin{enumerate}
    \item  If the interactive term of the Lagrangian $\lag_{int}$ (and consequently $L_{int}$) is invariant under the transformation
				\begin{equation}\label{su2transformation}\begin{split}
				\psi_1\rightarrow&\ct \psi_1-\st \psi_2\\
				\psi_2\rightarrow&\st \psi_1+\ct \psi_2
				\end{split}\end{equation}
	we can then write
	\be\label{conditionselfinteraction}
	\fnormal{\lag_{int}}=\fnormal{L_{int}}=\fnormal{\sum_i\bpsi_i\left[L_{int},\psi_i\right]}=0
	\ee
	leading to $w=0$.
    \item If $\lag_{int}$ is not invariant under (\ref{su2transformation}),
	we cannot push our analysis farther and we are unable to decide whether the pressure is zero or not,
	provided just with the tools here presented.
	It should be emphasized that other cancelation mechanisms might occur, leading to a full generalization of $w=0$
	for \textit{all} self-interactive cases, just like in the bosonic case.
	However, these mechanisms are not reproduced within our method.
\end{enumerate}

\subsection{Remarks on Interactive Theories}

To conclude the discussion of these examples, a few remarks are in order.
Throughout our analysis we assumed that the flavour vacuum was \textit{defined} by
\be\label{interactivefv}
\rfv\equiv G^{\dagger}_{\theta}\rmv
\ee
with $G_{\theta}$ the operator mapping flavour fields into massive fields, and vice versa,
and $\rmv$ being the massive ground state of the interactive theory.
The derivation of our results was purely formal and did not require any other knowledge of the theory.
Nonetheless, although it might look reasonable, the assumption (\ref{interactivefv}) remains a mere guess
in absence of a complete (either perturbative or non-perturbative) interactive theory.

An interactive theory is a rather different object than a free one, from a non-perturbative level.
In Section \ref{section:BVformalism}, we already mentioned that the usual Fock space \hilbert{F_0}
is not sufficient for fully describing the theory.
More generally we can say that in the framework of Second Quantization few
progresses on a coherent definition of the theory have been made so far, and
the explicit construction of physical states in interactive theories
still represents an open issue (\textit{cf} \cite{Borne:2002} and references therein).

Moreover, the familiar Perturbation Theory scheme,
in the formulation of Lehmann, Symanzik and Zimmermann \cite{Lehmann:1954rq},
is thought specifically for scattering processes and it might be unfit for describing
the \textit{flavour vacuum}. Since it relies on the assumption that particles
are free at early and late times,
all relevant quantities (scattering probabilities)
are expressed in terms of time ordered products of field acting on the vacuum of the free theory $\rmv$,
which is suppose to coincide with the true vacuum of theory at early and late times.
However, the features of the flavour vacuum are not expressed in these terms, \textit{i.e.}
as probabilities of having certain states at late times, given some initial conditions.

It follows that implementing BV formalism on interactive theories is not a trivial task and requires
very much care.
Such a generalization is not among the aims of the present work.
However, the purpose of this Section was to indicate a possible path for further developments of the formalism,
taking advantage of the method of calculation discussed so far.
Although an interactive theory might suffer from serious problems
when it comes to construct particle states, as above mentioned, we believe that
certain quantities, such as the equation of state of the flavour vacuum,
might not require an explicit expression of such a state.
Our analysis is valid under the assumption that (\ref{interactivefv}) holds,
irrespectively of a detailed knowledge of $\rmv$ or any particle states in
the interactive theory.
Therefore, we might expect to be able to get some features of the phenomenology of the flavour
vacuum, even though the underlying theory is not understood in full detail.
However, a dedicated analysis is in order to fully justify the use of (\ref{interactivefv}).

\section*{Conclusions}
Neutrino physics required in the last years a dedicated theoretical effort,
beyond the usual quantum field theoretical framework of scattering processes \cite{Beuthe:2001rc}.
Among other approaches, BV formalism is aimed to describe flavour states in a completely
non-perturbative way \cite{Blasone:1995zc}.
The approach fostered an intense discussion in last years \cite{Giunti:2003dg,Giunti:2006fr,Li:2006qt,Blasone:1999jb,Blasone:2005ae},
and it has not being accepted by the community as a whole.
However, we believe it represents a valid starting point
towards a complete and coherent treatment of non-perturbative aspects of flavour physics.

Besides providing a consistent relativistic generalization for
well-established non-relativistic neutrino oscillations formulae,
the formalism implies a non-trivial vacuum (the so called \textit{flavour vacuum}),
which has been regarded as a source of Dark Energy in various works \cite{Capolupo:2006et}.
Recently, the formalism has been used to describe the low energy limit of a quantum gravity model \cite{mavrosarkar},
leading to an implementation on a supersymmetric model \cite{Mavromatos:2010ni}.
Preliminary works have shown that BV formalism gives rise to a
novel mechanism of SUSY breaking \cite{Mavromatos:2010ni,Capolupo:2010ek}.
This work moves a step forth in the analysis of the supersymmetric flavour vacuum.

On one hand, we analyzed the phenomenology of the model of \cite{Mavromatos:2010ni,Capolupo:2010ek}
(a free Wess-Zumino with flavour mixing), arguing that the supersymmetric flavour vacuum
might consistently provide a source for both Dark Energy (thanks to the bosonic sector of the theory) and Dark Matter
(via the fermionic one).
At the moment, a quantitative comparison with available data is not yet possible,
due to the oversimplifications of the model (just
free fields labeled by only two flavours have been considered).
However, encouraging results come from a preliminary analysis here performed:
despite the huge difference of magnitudes of the involved parameters
(ranging from the Plank energy scale to neutrino masses,
from the density of Dark Matter today and its density at very early times),
the model seems to be capable of reproducing the hierarchy $(\Delta m^2)^2\sim \rho_\Lambda$,
in the assumption that all Dark energy density $\rho_\Lambda$ is due to the bosonic sector of the
flavour vacuum, and compatible with the hypothesis that the flavour vacuum
contributes to (at least a fraction of) Dark Matter.

On the other hand, we started developing new tools in order to test
the behaviour of the flavour vacuum in interactive theories.
We presented a novel method for evaluating relevant quantities connected with our problem,
which, not only facilitate the non-trivial calculations involved in the free model,
but also opens the way towards non-perturbative analyses of interactive theories.
As a concrete example of the advantages of the method, we first
showed how to reproduce in a few lines results of \cite{Mavromatos:2010ni,Capolupo:2010ek},
getting a deeper insight of known formulae. Furthermore, in order to test the potential
of the method, we used it to analyze the equation of state of the flavour vacuum in
self-interactive theories, generalizing results of the free theory for a wide class of
interactions, under reasonable assumptions.

The possibility here discussed that a source for both Dark Matter and Dark Energy might arise from
neutrino physics, whether it derives from new physics beyond the Standard Model or
from non-perturbative aspects of QFT, is quite attractive.
The model here presented needs to be understood and developed much further.
In particular, the following developments are in order:
\begin{itemize}
	\item more realistic theories need to be constructed and compared with observational data:
    three flavours, SM/MSSM interactions, evolution in time (possibly on a curved background) are essential ingredients;
    \item provided with a more realistic theory, the behaviour of the flavour vacuum on large distance scales
    must be examined in presence of matter, in order to be compared with phenomenological models of
    Dark Matter and Dark Energy;
	\item if we regard BV formalism as an effective description at low energy scales of the stringy model
    of \cite{mavrosarkar}, the gap between the macroscopic and microscopic description
    needs be reduced\footnote{An upcoming work of the authors of \cite{mavrosarkar}
    will start filling this gap.};
    \item other ways to prove experimentally the existence of the flavour vacuum must be found; besides
    its gravitational effects, it might play an active role in interactive theories and hence in scattering
    processes, which require a dedicated analysis;
\end{itemize}
Despite these and many other questions that remain open,
the models presented in this work suggest an intriguing possibility
for a deeper understanding of fundamental problems in cosmology. The promising results here discussed certainly motivate, we believe, further developments of the approach.


\begin{acknowledgments}
The author is thankful to N. E. Mavromatos and S. Sarkar for useful discussions.
\end{acknowledgments}
\bibliographystyle{plain}
\bibliography{biblio}

\begin{thebibliography}{10}

\bibitem{amendola}
L.~Amendola and S.~Tsujikawa.
\newblock {\em {Dark Energy: Theory and Observations}}.
\newblock Cambridge University Press, 2010.

\bibitem{Bertone:2004pz}
Gianfranco Bertone, Dan Hooper, and Joseph Silk.
\newblock {Particle dark matter: Evidence, candidates and constraints}.
\newblock {\em Phys. Rept.}, 405:279--390, 2005.

\bibitem{Beuthe:2001rc}
Mikael Beuthe.
\newblock {Oscillations of neutrinos and mesons in quantum field theory}.
\newblock {\em Phys. Rept.}, 375:105--218, 2003.

\bibitem{Bilenky:2010zza}
Samoil Bilenky.
\newblock {Introduction to the physics of massive and mixed neutrinos}.
\newblock {\em Lect. Notes Phys.}, 817:1--255, 2010.

\bibitem{Birrell:1982ix}
N.~D. Birrell and P.~C.~W. Davies.
\newblock {\em {QUANTUM FIELDS IN CURVED SPACE}}.
\newblock Cambridge, Uk: Univ. Pr. ( 1982) 340p.

\bibitem{Blasone:2004yh}
M.~Blasone, A.~Capolupo, S.~Capozziello, S.~Carloni, and Giuseppe Vitiello.
\newblock {Neutrino mixing contribution to the cosmological constant}.
\newblock {\em Phys. Lett.}, A323:182--189, 2004.

\bibitem{Blasone:2006ch}
M.~Blasone, A.~Capolupo, S.~Capozziello, and G.~Vitiello.
\newblock {Neutrino mixing and dark energy}.
\newblock {\em AIP Conf. Proc.}, 841:406--409, 2006.

\bibitem{Blasone:2007iq}
M.~Blasone, A.~Capolupo, S.~Capozziello, and G.~Vitiello.
\newblock {Cosmological effects of neutrino mixing}.
\newblock {\em AIP Conf. Proc.}, 957:185--188, 2007.

\bibitem{Blasone:2008rx}
M.~Blasone, A.~Capolupo, S.~Capozziello, and G.~Vitiello.
\newblock {A new perspective in the dark energy puzzle from particle mixing
  phenomenon}.
\newblock 2008.

\bibitem{Blasone:2007jm}
M.~Blasone, A.~Capolupo, S.~Capozziello, and G.~Vitiello.
\newblock {Neutrino mixing, flavor states and dark energy}.
\newblock {\em Nucl. Instrum. Meth.}, A588:272--275, 2008.

\bibitem{Blasone:2008ii}
M.~Blasone, A.~Capolupo, C.~R. Ji, and G.~Vitiello.
\newblock {On flavor violation for massive and mixed neutrinos}.
\newblock {\em Nucl. Phys. Proc. Suppl.}, 188:37--39, 2009.

\bibitem{Blasone:1995zc}
M.~Blasone and Giuseppe Vitiello.
\newblock {Quantum field theory of fermion mixing}.
\newblock {\em Ann. Phys.}, 244:283--311, 1995.

\bibitem{Blasone:2004hr}
Massimo Blasone, Antonio Capolupo, Salvatore Capozziello, Sante Carloni, and
  Giuseppe Vitiello.
\newblock {Neutrino mixing as a source for cosmological constant}.
\newblock {\em Braz. J. Phys.}, 35:455--461, 2005.

\bibitem{Blasone:2006jx}
Massimo Blasone, Antonio Capolupo, Chueng-Ryong Ji, and Giuseppe Vitiello.
\newblock {On flavor conservation in weak interaction decays involving mixed
  neutrinos}.
\newblock {\em Int. J. Mod. Phys.}, A25:4179--4194, 2010.

\bibitem{Blasone:2001du}
Massimo Blasone, Antonio Capolupo, Oreste Romei, and Giuseppe Vitiello.
\newblock {Quantum field theory of boson mixing}.
\newblock {\em Phys. Rev.}, D63:125015, 2001.

\bibitem{Blasone:2005ae}
Massimo Blasone, Antonio Capolupo, Francesco Terranova, and Giuseppe Vitiello.
\newblock {Lepton charge and neutrino mixing in pion decay processes}.
\newblock {\em Phys.Rev.}, D72:013003, 2005.

\bibitem{Blasone:2001sr}
Massimo Blasone, Antonio Capolupo, and Giuseppe Vitiello.
\newblock {Comment on 'Remarks on flavor-neutrino propagators and oscillation
  formulae'}.
\newblock 2001.

\bibitem{Blasone:2002jv}
Massimo Blasone, Antonio Capolupo, and Giuseppe Vitiello.
\newblock {Quantum field theory of three flavor neutrino mixing and
  oscillations with CP violation}.
\newblock {\em Phys. Rev.}, D66:025033, 2002.

\bibitem{Blasone:1998hf}
Massimo Blasone, Peter~A. Henning, and Giuseppe Vitiello.
\newblock {The exact formula for neutrino oscillations}.
\newblock {\em Phys. Lett.}, B451:140--145, 1999.

\bibitem{Blasone:1999jb}
Massimo Blasone and Giuseppe Vitiello.
\newblock {Remarks on the neutrino oscillation formula}.
\newblock {\em Phys. Rev.}, D60:111302, 1999.

\bibitem{Borne:2002}
T.~Borne, G.~Lochak, and H.~Stumpf.
\newblock {\em {Nonperturbative Quantum Field Theory and the Structure of
  Matter}}.
\newblock Fundamental theories of physics. Kluwer Academic Pub, 2001.

\bibitem{calzetta2008nonequilibrium}
E.A. Calzetta and B.L. Hu.
\newblock {\em Nonequilibrium quantum field theory}.
\newblock Cambridge monographs on mathematical physics. Cambridge University
  Press, 2008.

\bibitem{Capolupo:2007hy}
A.~Capolupo, S.~Capozziello, and G.~Vitiello.
\newblock {Dark energy, cosmological constant and neutrino mixing}.
\newblock {\em Int. J. Mod. Phys.}, A23:4979--4990, 2008.

\bibitem{Capolupo:2008rz}
A.~Capolupo, S.~Capozziello, and G.~Vitiello.
\newblock {Dark energy and particle mixing}.
\newblock {\em Phys. Lett.}, A373:601--610, 2009.

\bibitem{Capolupo:2006et}
A.~Capolupo, Salvatore Capozziello, and G.~Vitiello.
\newblock {Neutrino mixing as a source of dark energy}.
\newblock {\em Phys. Lett.}, A363:53--56, 2007.

\bibitem{Capolupo:2004av}
Antonio Capolupo.
\newblock {Aspects of particle mixing in quantum field theory}.
\newblock 2004.

\bibitem{Capolupo:2006re}
Antonio Capolupo, Salvatore Capozziello, and Giuseppe Vitiello.
\newblock {Dark energy induced by neutrino mixing}.
\newblock {\em J. Phys. Conf. Ser.}, 67:012032, 2007.

\bibitem{Capolupo:2010ek}
Antonio Capolupo, Marco Di~Mauro, and Alfredo Iorio.
\newblock {Mixing-induced Spontaneous Supersymmetry Breaking}.
\newblock {\em Phys.Lett.}, A375:3415--3418, 2011.

\bibitem{Ellis:2004ay}
John~R. Ellis, Nikolaos~E. Mavromatos, and Michael Westmuckett.
\newblock {A supersymmetric D-brane model of space-time foam}.
\newblock {\em Phys. Rev.}, D70:044036, 2004.

\bibitem{Ellis:2005ib}
John~R. Ellis, Nikolaos~E. Mavromatos, and Michael Westmuckett.
\newblock {Potentials between D-branes in a supersymmetric model of space-time
  foam}.
\newblock {\em Phys. Rev.}, D71:106006, 2005.

\bibitem{Fujii:1998xa}
Kanji Fujii, Chikage Habe, and Tetsuo Yabuki.
\newblock {Note on the field theory of neutrino mixing}.
\newblock {\em Phys. Rev.}, D59:113003, 1999.

\bibitem{Fujii:2001zv}
Kanji Fujii, Chikage Habe, and Tetsuo Yabuki.
\newblock {Remarks on flavor-neutrino propagators and oscillation formulae}.
\newblock {\em Phys. Rev.}, D64:013011, 2001.

\bibitem{Giunti:2007ry}
C.~Giunti and C.W. Kim.
\newblock {\em Fundamentals of neutrino physics and astrophysics}.

\bibitem{Giunti:2003dg}
Carlo Giunti.
\newblock {Fock states of flavor neutrinos are unphysical}.
\newblock {\em Eur. Phys. J.}, C39:377--382, 2005.

\bibitem{Giunti:2006fr}
Carlo Giunti.
\newblock {Neutrino Flavor States and Oscillations}.
\newblock {\em J.Phys.G}, G34:R93--R109, 2007.

\bibitem{Gribov:1968kq}
V.~N. Gribov and B.~Pontecorvo.
\newblock {Neutrino astronomy and lepton charge}.
\newblock {\em Phys. Lett.}, B28:493, 1969.

\bibitem{Gurwich:2010gb}
Ilya Gurwich.
\newblock {Natural Neutrino Dark Energy}.
\newblock 2010.

\bibitem{Haag:1955ev}
R.~Haag.
\newblock {On quantum field theories}.
\newblock {\em Kong. Dan. Vid. Sel. Mat. Fys. Med.}, 29N12:1--37, 1955.

\bibitem{Hannabuss:2000hy}
K.~C. Hannabuss and D.~C. Latimer.
\newblock {The quantum field theory of fermion mixing}.
\newblock {\em J. Phys.}, A33:1369--1373, 2000.

\bibitem{Hawking:1974sw}
S.~W. Hawking.
\newblock {Particle Creation by Black Holes}.
\newblock {\em Commun. Math. Phys.}, 43:199--220, 1975.

\bibitem{Ji:2001yd}
Chueng-Ryong Ji and Yuriy Mishchenko.
\newblock {Nonperturbative vacuum effect in the quantum field theory of meson
  mixing}.
\newblock {\em Phys. Rev.}, D64:076004, 2001.

\bibitem{Ji:2002tx}
Chueng-Ryong Ji and Yuriy Mishchenko.
\newblock {The general theory of quantum field mixing}.
\newblock {\em Phys. Rev.}, D65:096015, 2002.

\bibitem{Jungman:1995df}
Gerard Jungman, Marc Kamionkowski, and Kim Griest.
\newblock {Supersymmetric dark matter}.
\newblock {\em Phys. Rept.}, 267:195--373, 1996.

\bibitem{Komatsu:2010fb}
E.~Komatsu et~al.
\newblock {Seven-Year Wilkinson Microwave Anisotropy Probe (WMAP) Observations:
  Cosmological Interpretation}.
\newblock {\em Astrophys. J. Suppl.}, 192:18, 2011.

\bibitem{Lehmann:1954rq}
H.~Lehmann, K.~Symanzik, and W.~Zimmermann.
\newblock {On the formulation of quantized field theories}.
\newblock {\em Nuovo Cim.}, 1:205--225, 1955.

\bibitem{Li:2006qt}
Y.F. Li and Q.Y. Liu.
\newblock {A Paradox on quantum field theory of neutrino mixing and
  oscillations}.
\newblock {\em JHEP}, 0610:048, 2006.

\bibitem{Liddle:2009zz}
Andrew Liddle.
\newblock {An introduction to modern cosmology}.
\newblock Weinheim, Germany: Wiley-VCH (2009) 201 p.

\bibitem{Maki:1962mu}
Ziro Maki, Masami Nakagawa, and Shoichi Sakata.
\newblock {Remarks on the unified model of elementary particles}.
\newblock {\em Prog. Theor. Phys.}, 28:870--880, 1962.

\bibitem{Mavromatos:2006yy}
N.~E. Mavromatos and Sarben Sarkar.
\newblock {Methods of approaching decoherence in the flavour sector due to
  space-time foam}.
\newblock {\em Phys. Rev.}, D74:036007, 2006.

\bibitem{Mavromatos:2005bu}
Nick Mavromatos and Sarben Sarkar.
\newblock {Liouville decoherence in a model of flavour oscillations in the
  presence of dark energy}.
\newblock {\em Phys. Rev.}, D72:065016, 2005.

\bibitem{mavrosarkar}
Nick~E. Mavromatos and Sarben Sarkar.
\newblock {Towards a microscopic construction of flavour vacua from a space-time foam model}.
\newblock {\em New J.Phys.}, 10:073009, 2008.

\bibitem{Mavromatos:2008bz}
Nick~E. Mavromatos and Sarben Sarkar.
\newblock {Non-extensive statistics in stringy space-time foam models}.
\newblock {\em Phys. Rev.}, D79:104015, 2009.

\bibitem{Mavromatos:2009rf}
Nick~E. Mavromatos, Sarben Sarkar, and Walter Tarantino.
\newblock {Flavour Condensates in Brane Models and Dark Energy}.
\newblock {\em Phys.Rev.}, D80:084046, 2009.

\bibitem{Mavromatos:2010ni}
Nick~E. Mavromatos, Sarben Sarkar, and Walter Tarantino.
\newblock {D-foam-induced Flavour-Condensate-induced Breaking of Supersymmetry in Free Wess-Zumino Fluids}.
\newblock {\em Phys.Rev.}, D84:044050, 2011.

\bibitem{Peskin:1995ev}
Michael~E. Peskin and Daniel~V. Schroeder.
\newblock {\em {An Introduction to quantum field theory}}.
\newblock Reading, USA: Addison-Wesley (1995) 842 p.

\bibitem{Pontecorvo:1957qd}
B.~Pontecorvo.
\newblock {Inverse beta processes and nonconservation of lepton charge}.
\newblock {\em Sov. Phys. JETP}, 7:172--173, 1958.

\bibitem{Pontecorvo:1967fh}
B.~Pontecorvo.
\newblock {Neutrino experiments and the question of leptonic-charge
  conservation}.
\newblock {\em Sov. Phys. JETP}, 26:984--988, 1968.

\bibitem{Streater:1989vi}
R.~F. Streater and A.~S. Wightman.
\newblock {\em {PCT, spin and statistics, and all that}}.
\newblock Redwood City, USA: Addison-Wesley (1989) 207 p. (Advanced book
  classics).

\bibitem{Strumia:2006db}
Alessandro Strumia and Francesco Vissani.
\newblock {Neutrino masses and mixings and..}
\newblock 2006.

\bibitem{Umezawa:1982nv}
H.~Umezawa, H.~Matsumoto, and M.~Tachiki.
\newblock {\em {THERMO FIELD DYNAMICS AND CONDENSED STATES}}.
\newblock Amsterdam, Netherlands: North-holland ( 1982) 591p.

\bibitem{Unruh:1976db}
W.~G. Unruh.
\newblock {Notes on black hole evaporation}.
\newblock {\em Phys. Rev.}, D14:870, 1976.

\end{thebibliography}
\end{document}